# Coherent phonon dynamics in spatially separated graphene mechanical resonators

Zhuo-Zhi Zhang[1,2,†], Xiang-Xiang Song[1,2,†,*], Gang Luo[1,2,†], Zi-Jia Su[1,2], Kai-Long Wang[1,2], Gang Cao[1,2], Hai-Ou Li[1,2], Ming Xiao[1,2], Guang-Can Guo[1,2], Lin Tian[3,*], Guang-Wei Deng[1,2,*], and Guo-Ping Guo[1,2,4,*]

1. CAS Key Laboratory of Quantum Information, University of Science and Technology of China, Hefei, Anhui 230026, China
2. CAS Center for Excellence in Quantum Information and Quantum Physics, University of Science and Technology of China, Hefei, Anhui 230026, China
3. School of Nature Sciences, University of California, Merced, California 95343, USA
4. Quantum Chip Lab, Origin Quantum Computing Company Limited, Hefei, Anhui 230088, China

† Z.-Z. Zhang, X.-X. Song, and G. Luo contributed equally to this work.
* Correspondence and requests for materials should be addressed to X.-X. Song (songxx90@ustc.edu.cn), L. T. (ltian@ucmerced.edu), G.-W. Deng (gwdeng@ustc.edu.cn), or G.-P. G. (gpguo@ustc.edu.cn ).

## Abstract

Vibrational modes in mechanical resonators provide a promising candidate to interface and manipulate classical and quantum information. The observation of coherent dynamics between distant mechanical resonators can be a key step towards scalable phonon-based applications. Here we report tunable coherent phonon dynamics with an architecture comprising three graphene mechanical resonators coupled in series, where all resonators can be manipulated by electrical signals on control gates. We demonstrate coherent Rabi oscillations between spatially separated resonators indirectly coupled via an intermediate resonator serving as a phonon cavity. The Rabi frequency fits well with the microwave burst power on the control gate. We also observe Ramsey interference, where the oscillation frequency corresponds to the indirect coupling strength between these resonators. Such coherent processes indicate that information encoded in vibrational modes can be transferred and stored between spatially separated resonators, which can open the venue of on-demand phonon-based information processing.

**Significance Statement**

The rapid development of nanotechnologies enables the possibility to use long-lifetime vibrational phonon modes for classical and quantum information processing. Despite continuous efforts in the past decade, tunable phonon-mediated transferring and processing of information between distant phonon modes remains technically challenging. Taking advantage of the excellent electronic and mechanical properties of graphene, we are able to realize electrically tunable coherent phonon dynamics between spatially separated graphene mechanical resonators. The gate-controllable resonant frequencies, coupling strengths, and coherent phonon dynamics indicate that information encoded in vibrational modes can be stored, transferred, and manipulated between spatially separated resonators. Our results not only demonstrate the transfer of information between phonons, but also provide the building blocks towards scalable phonon-based information processing.

# Introduction

Over the past decade, one of the major achievements in nanotechnology was the realization of phonon state storage and transfer (1-9). Scalable phonon-based classical (10) and quantum (6, 7) information processing requires manipulation of highly coherent phonon modes with tunable coupling. Efforts towards phonon state manipulation have been made in different systems, such as the dynamical control of phonon states between different modes in a single mechanical resonator (10-12) or between modes in neighbouring mechanical resonators (13), the observation of static phonon states of spatially separated mechanical resonators with indirect coupling (14), and routing of microwave phonons using phonon waveguide (15). Theoretical works also proposed that couplings between two spatially separated mechanical oscillators via the exchange of virtual photon pairs can be achieved (16). However, observation of tunable coherent dynamics of indirectly coupled phonon modes remains a challenge, due to the requirement on the combination of strong coupling, long coherence time, and tunability of phonon modes.

Graphene-based mechanical resonators with high quality factors and excellent electrical control provides us with a promising platform for tunable phonon manipulations (17-21). Here we use high-quality graphene-based mechanical resonators to demonstrate coherent phonon dynamics. We report the observation of coherent Rabi and Ramsey oscillations (22, 23) between spatially separated mechanical resonators in the classical regime via Raman-like indirect coupling (14). Our results indicate that phonon coherence can be maintained and controlled over distant resonators, which will shed light on scalable phonon-based information processing.

## Results

**Sample characterization**

As shown in Figs. 1**a** and **b**, a graphene ribbon with a width of ~2.2 μm and ~7 layers is suspended over three trenches (2 μm in width, 200 nm in depth) between four contact electrodes. This structure defines three electro-mechanical resonators, labelled as $R_1$, $R_2$, and $R_3$ in Fig. 1**a**, respectively. All measurements were carried out in a dilution refrigerator at a base temperature of ~ 10 mK and a pressure below $10^{-7}$ Torr. Similar structure was used to realize indirect coupling between mechanical resonators via a Raman-like process with static presentations (14). Here optimising the measurements setup and the quality of the device allows us to employ this system to explore tunable coherent phonon dynamics in spatially separated mechanical resonators.

We characterize each resonator separately using the one-source frequency modulation technique (24), where a frequency-modulated (FM) signal, $V^{\mathrm{FM}}(t) = V^{\mathrm{AC}} \cos[2\pi f_\mathrm{d} t + \frac{f_\Delta}{f_\mathrm{L}} \sin(2\pi f_\mathrm{L} t)]$, is applied at a selected contact, and the mixing current $I_{\mathrm{mix}}$ at frequency $f_\mathrm{L}$ is recorded by a lock-in amplifier at another contact, as illustrated in Fig. 1**b**. We find that the relaxation rate of the resonators shows strong dependency on the driving amplitude. At a low driving amplitude, ~ −60 dBm, we extract the relaxation rate to be $\gamma_1/2\pi = 1.20$ kHz from the spectrum of the mixing current $I_{\mathrm{mix}}$ by fitting (24), as shown in Fig. 1**c** and Fig. S2, which corresponds to a quality factor of $Q_1 = f_{\mathrm{m1}}/\gamma_1 \times 2\pi \sim 111{,}000$ for resonator $R_1$. Comparing to our previous work (14), observed quality factor is increased mainly due to that frequency-modulation technique allows for lower probing power (25). At a stronger driving power of $\sim -45$ dBm, the relaxation rate becomes $\gamma_1/2\pi \sim 300$ kHz, corresponding to a much lower quality factor of $\sim 300$. The relaxation rate increases significantly with increasing driving power due to nonlinear damping, which is similar to previous reports in nanomechanical systems (26, 27). Note that we cannot estimate the magnitude of frequency fluctuations (28) and eliminate its influence in present situation, what we obtained here is an upper bound of the relaxation rate. Here the resonant frequencies of all three resonators ($f_{mi}$, i =1, 2, and 3) can be tuned by adjusting the corresponding gate voltages $V_{gi}$ (i =1, 2, 3). Details of the characterization of the resonators can be found in the SI Appendix (Figs. S1 and S2).

**Strongly coupled mechanical resonators**

To characterize the coupling between the resonators $R_1$ and $R_2$, we fix the frequency $f_{m2}$ of $R_2$ (i.e., fix the gate voltage on gate $g2$) and tune $f_{m1}$ by adjusting $V_{g1}$. When $f_{m1}$ approaches $f_{m2}$, an avoided level crossing appears in Fig. 1**d**. By measuring the frequency splitting, the coupling strength between neighbouring resonators was obtained with $\Omega_{12}/2\pi \sim 11.5$ MHz. Here the coupling between the neighbouring resonators $R_1$ and $R_2$ leads to a normal mode that is spatially distributed over the range of $R_1$ and $R_2$. Similarly, we can measure the coupling strength between the neighbouring resonators $R_2$ and $R_3$, $\Omega_{23}/2\pi$, to be $\sim 9.0$ MHz, as shown in Fig. S3. Since both $R_1$ and $R_3$ are neighbouring to $R_2$, avoided level crossing between $R_1$ and $R_3$ mediated by $R_2$ is observed, which can be translated to an effective coupling between the non-neighbouring resonators ($R_1$ and $R_3$). The indirect coupling strength $\Omega_{13}/2\pi$ between the spatially separated resonators $R_1$ and $R_3$ is shown in Fig. 1e. The Raman-like effective coupling (14) $\Omega_{13}/2\pi$ decreases when the detuning $\Delta_{23}$ between $f_{m2}$ and $f_{m3}$ increases, as shown in Fig. 1f. With $\Omega_{13} \gg \gamma_1, \gamma_3$, the system is well in the strong-coupling regime.

Cooperativity is a key parameter to characterize the coherence and the potential for coherent manipulation of coupled modes. At low driving power $\sim -60$ dBm, we can calculate the zero-temperature cooperativities with $C_{12} = \Omega_{12}^2/\gamma_1\gamma_2$ for resonators $R_1$ and $R_2$ reaching $9.5 \times 10^7$ and $C_{23} = \Omega_{23}^2/\gamma_2\gamma_3$ for resonators $R_2$ and $R_3$ reaching $9.6 \times 10^7$ (see the SI Appendix Fig. S3 for details). More importantly, with effective coupling strength, $\Omega_{13}/2\pi \sim 3.3$ MHz, between the non-neighbouring resonators $R_1$ and $R_3$, the zero-temperature cooperativity $C_{13} = \Omega_{13}^2/\gamma_1\gamma_3$ between these spatially separated modes exceeds $1.2 \times 10^7$, as shown in Figs. S4 and S5. Such giant cooperativity is several orders of magnitude larger than previously reported results in coupled-mechanical systems (10, 11, 13, 14, 18, 19, 29-31). While for large driving power ($\sim -45$ dBm), where we conduct measurement on the coherent oscillations (see Figs. 2 and 3), the increased relaxation rate ($\sim 300$ kHz) leads to zero-temperature cooperativities of $C_{12}$, $C_{23} \sim 1000$ and $C_{13} \sim 100$. As mentioned above, the measured relaxation rate includes the effect of frequency fluctuations, which indicates that the actual cooperativity can be much larger. It is worthy to note that when considering thermal occupation (with average thermal phonon number $\bar{n}_{th} = 1/(e^{\hbar\omega/k_B T} - 1) \sim 1.61$, where $k_B$ is Boltzmann constant, $T$ is around 10 mK), the finite-temperature cooperativity $C_{13}^{th} = C_{13}/\bar{n}_1^{th}\bar{n}_3^{th}$ is still sufficiently high for observing coherent phonon dynamics shown in Figs. 2 and 3.

**Rabi oscillations between spatially separated resonators**

The high cooperativity between resonators $R_1$ and $R_3$ is the basis for high fidelity state transfer within these spatially separated modes. To demonstrate this, we perform time-domain experiments to study such coherent oscillations. First, we bias the system at $V_{g1} = 18$ V, $V_{g2} = 17.5$ V, and $V_{g3} = 22$ V. At this condition, the resonator $R_1$ and $R_3$ are degenerate, as indicated by the anti-crossing between these modes. We initialize the coupled system to be excited at the lower eigenmode, which is labelled as $A$ in Fig. 2**a**, with $A$ being the symmetric superposition of $R_1$ and $R_3$. We then apply a microwave burst at the frequency $f_{MW} = \Omega_{13}/2\pi = 3.11$ MHz on gate $g1$. The measured mixing current demonstrates Rabi oscillations between the lower eigenmode $A$ and the upper eigenmode $B$, as a function of the duration of the burst. In analogy to a two-level system, the dynamics of the phonon excitation can be illustrated with the Bloch sphere of theses eigenmodes (Fig. 2b), where eigenmode $A$ is the south pole of the sphere, and eigenmode $B$ is the north pole (10, 23). Under the microwave burst, the system rotates around the $x$ axis of the Bloch sphere and exhibits a Rabi oscillation. The population of eigenmode $A$ oscillates with the duration $\tau$ of the burst, which generates coherent oscillation in the mixing current $I_{mix}$ detected at the frequency of the eigenmode $A$. The mixing current versus the microwave duration $\tau$ at different pulse amplitude $V_{pp}$ (peak to peak) is plotted in Figs. 2**c** and **d**. The time dependence of $I_{mix}$ can be fitted with a damped cosinusoidal function, $Ae^{-\tau/T_{Rabi}}\cos(\Omega_{Rabi}\tau)$,

with damping time $T_{\text{Rabi}}$ and Rabi frequency $\Omega_{\text{Rabi}}/2\pi$. In the upper panel of Fig. 2**c**, the Rabi frequency is ~ 0.25 MHz with damping time $T_{\text{Rabi}}$ ~ 11.6 μs, with burst amplitude $V_{\text{pp}} = 0.60$ V. In Fig. 2**e**, the measured Rabi frequency is plotted versus the voltage amplitude $V_{\text{pp}}$ and shows a linear dependence on the amplitude, in agreement with our analysis (32). Details of eigenmodes $A$ and $B$ and their dynamics are presented in SI Appendix S7.

**Ramsey interference between spatially separated resonators**

Next, we study the precession dynamics between resonators $R_1$ and $R_3$ in a Ramsey interference experiment. The phonon excitation is initially prepared in eigenmode $A$, similar to that of the Rabi oscillation experiment. In the beginning of each cycle, a $\pi/2$ pulse ($V_{\text{pp}} = 0.60$ V and pulse duration ~ 1.1 μs) around the *x* axis is applied to this system, which rotates the state to the *x-y* plane in the Bloch sphere. This system then undergoes a free evolution of duration $\tau_{\text{wait}}$. During the free evolution, the system rotates around the *z* axis governed by $\Omega_{13}/2\pi$. A second $\pi/2$ pulse in the *x* axis is then applied to rotate the state. This process is illustrated in Fig. 3**a**. We measure the mixing current at the frequency of eigenmode $A$. The average mixing current versus $\tau_{\text{wait}}$ is shown in Fig. 3**b**, together with a damped cosinusoidal function $Ae^{-\tau/T_{\text{Ramsey}}}\cos(\Omega_{\text{Ramsey}}\tau)$ with fitting parameters $T_{\text{Ramsey}}$ and $\Omega_{\text{Ramsey}}$. Our result gives that $T_{\text{Ramsey}}$ ~ 9.5 μs and $\Omega_{\text{Ramsey}}/2\pi$ ~ 3.12 MHz, which is consistent with $\Omega_{13}/2\pi = 3.11$ MHz obtained from Fig. 2**a**. Furthermore, we adjust the relative phase between the two $\pi/2$ pulses in the Ramsey sequence from $-\pi$ to $\pi$ to change the rotation axis (33, 34). In Fig. 3**c,** the mixing current is plotted versus the angle of the rotation axis, $\Delta\varphi$, as well as the waiting time, and exhibits continuous variation of the excitation population in the $A$ mode.

**Rabi and Ramsey oscillations at different coupling strengths**

The highly tunable coupling strength between resonators $R_1$ and $R_3$ offers promising prospect for phonon state manipulation. We investigate the coherent dynamics under different coupling strength $\Omega_{13}/2\pi$, which can be tuned by varying the detuning between $R_1$ ($R_3$) and $R_2$. Our results are presented in Figs. 4**a-c** for Rabi oscillations and in Figs. 4**d-f** for Ramsey interference. The oscillation frequencies ($\Omega_{\text{Rabi}}/2\pi$ and $\Omega_{\text{Ramsey}}/2\pi$, respectively), and damping times ($T_{\text{Rabi}}$ and $T_{\text{Ramsey}}$, respectively) are extracted from the measured data. In Figs. 4**d** and **h**, we plot the damping times as well as an effective quality factor, defined as the damping time multiplies the corresponding oscillation frequency for the Rabi oscillations or the Ramsey interference. With our parameters, $\Omega_{\text{Rabi}}/2\pi$ remains almost the same ($\sim 0.25$ MHz) for different coupling strength $\Omega_{13}/2\pi$, indicating that the coupling between the microwave burst to the mechanical resonators stays almost the same as the effective coupling varies. The Ramsey frequency $\Omega_{\text{Ramsey}}/2\pi$ remains consistent with $\Omega_{13}/2\pi$. Meanwhile, $T_{\text{Rabi}}$ remains at $\sim 11$ μs, and $T_{\text{Ramsey}}$ varies from 5 μs to 10 μs, all having the same order of magnitude over large range of effective coupling. Combining with the results shown in Fig. 2**e**, we can adjust oscillation frequencies $\Omega_{\text{Rabi}}/2\pi$ and $\Omega_{\text{Ramsey}}/2\pi$ by electrical tuning, which demonstrates tunable phonon dynamics in spatially separated graphene mechanical resonators. The coherent oscillation experiments were taken at driving power of $\sim -45$ dBm. Under this driving power, the relaxation rate of each modes is ~300 kHz, leading to the damping time in the coherent oscillation experiments $T_{\text{Rabi}} = T_{\text{Ramsey}} \sim 2/\gamma_i$ to be ~1 μs (see SI Appendix S7 and S8), which is comparable with $T_{\text{Rabi}}$ and $T_{\text{Ramsey}}$ obtained from the experiment.

**Discussions**

It is worth noting that at low driving power with a quality factor up to $\sim 10^5$, the zero-temperature cooperativity between the resonators can reach $10^7$. The measured damping times could be improved by optimizing the measurement setup to lower driving power to decrease the influence of nonlinear damping effects (25) in the future. Moreover, frequency stabilization such as feedback technique (35) can be utilized to suppress frequency fluctuations (28). Our results provide a promising platform for coherent computation using controllable vibrational phonons. Meanwhile, taking advantages of mechanical resonators being an outstanding interface between different physical systems, such as quantum dots engineered in graphene flakes (21), our architecture can be applied to transferring and storage information between various systems. Although the present system is still a classical one, the phonon occupation number at undriven state is ~2 (given by $\bar{n}_{th} = 1/(e^{\hbar\omega/k_B T} - 1) \sim 1.61$). Using sideband cooling, advanced measurement setup, or resonators with higher frequency (36), our architecture can be pushed into quantum regime.

In conclusion, we studied coherent phonon dynamics between spatially separated graphene mechanical resonators. The giant cooperativity in our sample enables the study of such coherent oscillations and the fitting of the damping times. Our demonstration of coherent dynamics between mechanical modes with indirect interaction and high tunability is a key step towards on-demand state transfer and manipulation (6, 7, 33, 37) in an all-phonon platform. With the capabilities of system integrations, our architecture can be extended to large scale for phonon-based computing and long-distance information exchange in mechanical systems.

## Materials and Methods

### Sample fabrication

A layer of $SiN_x$ (50 nm) is deposited via low pressure chemical vapour deposition (LPCVD) on the silicon oxide layer, which covers a highly resistive silicon wafer. After electron beam lithography (EBL), three parallel trenches are etched through the $SiN_x$ layer by reactive ion etching, followed by dipping in buffered oxide etchant. The total etching depth is 200 nm. The widths of the trenches are designed as 2 µm. After a second EBL, 5 nm titanium and 20 nm gold are evaporated onto the wafer. All the contacts are 1.5 µm in widths. Finally, the graphene ribbon, exfoliated on a polydimethylsiloxane (PDMS) stamp, is aligned and transferred above the trenches (38).

### Measurement setup

For data acquisition in Figs. 1d-f and Fig. 2a, a frequency-modulated (FM) signal is generated by a Keysight E8257D microwave generator and applied to the Source of the centre resonator $R_2$, with deviation frequency $f_\Delta = 100$ kHz, modulation frequency $f_L = 1.33$ kHz, a Stanford Research SR830 lock-in amplifier is applied to read $I_{mix}$ with frequency $f_L$ at the Drain of $R_2$. For quality factor acquisition, $f_\Delta = 1$ kHz, $f_\mathrm{L} = 233$ Hz. In Rabi and Ramsey oscillation experiments, the driving FM signal is constantly applied at the frequency of the lower state of the superposition state, i.e., eigenmode $A$ in Fig. 2a. Additional driving pulse is generated by a Tektronix AWG7082C arbitrary waveform generator and applied to the gate electrode of $R_1$. Here, the lock-in amplifier is set at a time constant of 300 ms with frequency $f_\mathrm{L} = 1.33$ kHz. At each pulse parameter, $I_{mix}$ is acquired for over 40 times. The repetition rate of the waveform is 20 ~ 50 kHz in different experiments. Hence the total average times is over $2.4 \times 10^5 \sim 6 \times 10^5$. Here, the measured $I_{mix}$, which is averaged over many pulse cycles, can be recognized as the occupation of eigenmode $A$.

## Author Contributions

Z.-Z.Z., G.L., and K.-L.W. fabricated the device, Z.-Z.Z., X.-X.S., and G.-W.D. performed the measurements. L.T., X.-X.S., Z.-Z.Z., G.-W.D., and Z.-J.S. analysed the data and developed the theoretical analysis. H.-O.L., G.C., M.X., and G.-C.G. supported the fabrication and measurement. G.-P.G., X.-X.S., and G.-W.D. planned the project. Z.-Z.Z., X.-X.S., G.-W. D., and L.T. wrote the manuscript with input from all the other authors.

## Data availability

All data support the findings of this study are available within the article and/or the supporting information, with corresponding raw data deposited at Zenodo (39).

## Acknowledgements

The authors would like to thank J. Moser for fruitful discussions in sample fabrication and measurement techniques. This work was supported by the National Key Research and Development Program of China (Grant No. 2016YFA0301700), the National Natural Science Foundation of China (Grant Nos. 11625419, 61704164, 61674132, 11674300, 11575172, 61904171, and 11904351), China Postdoctoral Science Foundation (Grant Nos. BX20180295, 2018M640586), and the Anhui Initiative in Quantum Information Technologies (Grant No. AHY080000). L.T. is supported by the National Science Foundation under Award No. PHY-1720501 and the University of California Multicampus-National Lab Collaborative Research and Training under Award No. LFR-17-477237. This work was partially carried out at the University of Science and Technology of China Center for Micro and Nanoscale Research and Fabrication.

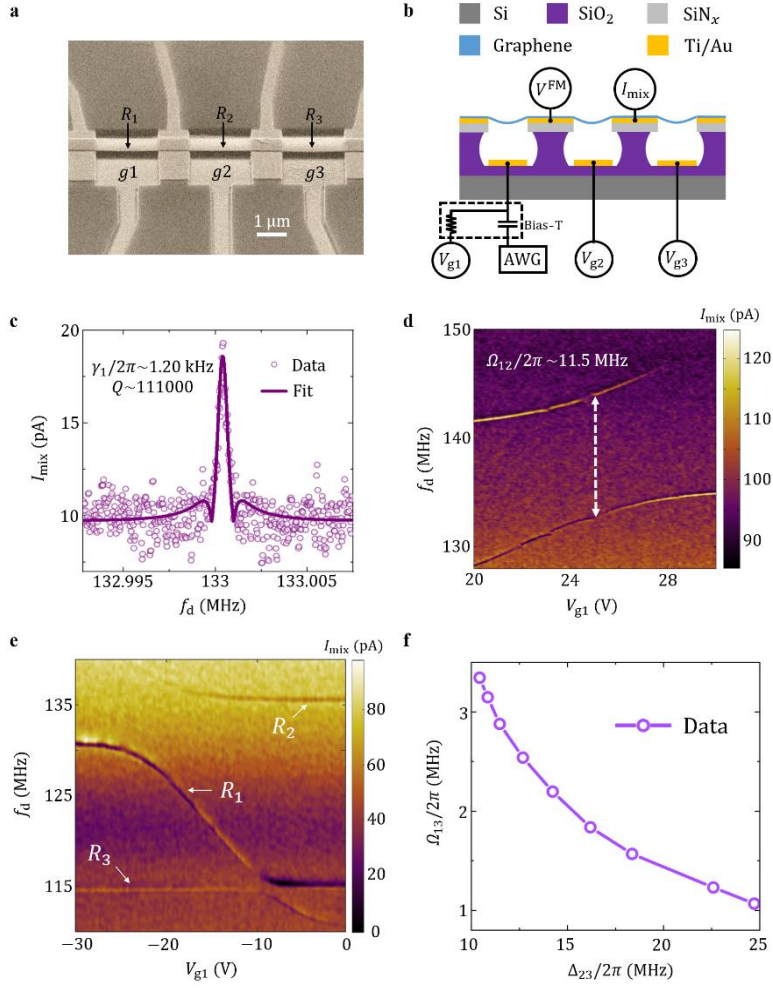

**Figure 1 | Sample structure and mode spectra of the resonators. a.** Scanning electron microscope image of a typical sample with a tilted view angle. Four contacts divide a suspended graphene ribbon into three mechanical resonators, $R_1$, $R_2$, and $R_3$, in a linear chain. **b.** Schematic sample structure and measurement setup. A frequency-modulated microwave signal $V^{FM}$ is applied through one contact of the centre resonator, and a lock-in amplifier detects the mixing current $I_{mix}$ at the modulation frequency of the inputted FM signal at another contact. The frequencies of the resonators are tuned by dc voltages on the gates, respectively. An arbitrary waveform generator (AWG) is connected to gate $g1$ to provide additional burst signals for the coherent oscillations. **c.** The mixing current as a function of the driving frequency $f_d$

at $V_{g1}$ = 22 V. The measured linewidth of resonator $R_1$ is $\gamma_1/2\pi$ ~1.20 kHz. Similarly, the linewidth of $R_2$ is $\gamma_2/2\pi$ ~1.16 kHz and the linewidth of $R_3$ is $\gamma_3/2\pi$ ~ 0.73 kHz (see SI Appendix Fig. S2) at a driving power of ~ −60 dBm. **d.** The mixing current spectrum of resonators $R_1$ and $R_2$. Coupling strength as large as 11.5 MHz is observed. Here $V_{g2}$ = 17 V and $V_{g3}$ = 0 V. **e.** Spectrum of all three resonators. Here $R_2$ is far off-resonance from $R_3$ with a detuning $\Delta_{23}/2\pi$ ~ 20 MHz, with $V_{g2}$ = 14 V, and $V_{g3}$ = 15 V. The dc voltage $V_{g1}$ is scanned over a wide range, to tune the resonant frequency of $R_1$ to cross the frequencies $f_{m2}$ and $f_{m3}$. A large avoided level crossing is observed when $f_{m1}$ approaches $f_{m2}$. A smaller energy splitting is observed when $f_{m1}$ approaches $f_{m3}$ due to the indirect coupling. **f.** The measured indirect coupling strength $\Omega_{13}/2\pi$ between $R_1$ and $R_3$ as a function of $\Delta_{23}/2\pi$. **d** and **e** are obtained at a driving power of ~ −40 dBm to have a better resolution.

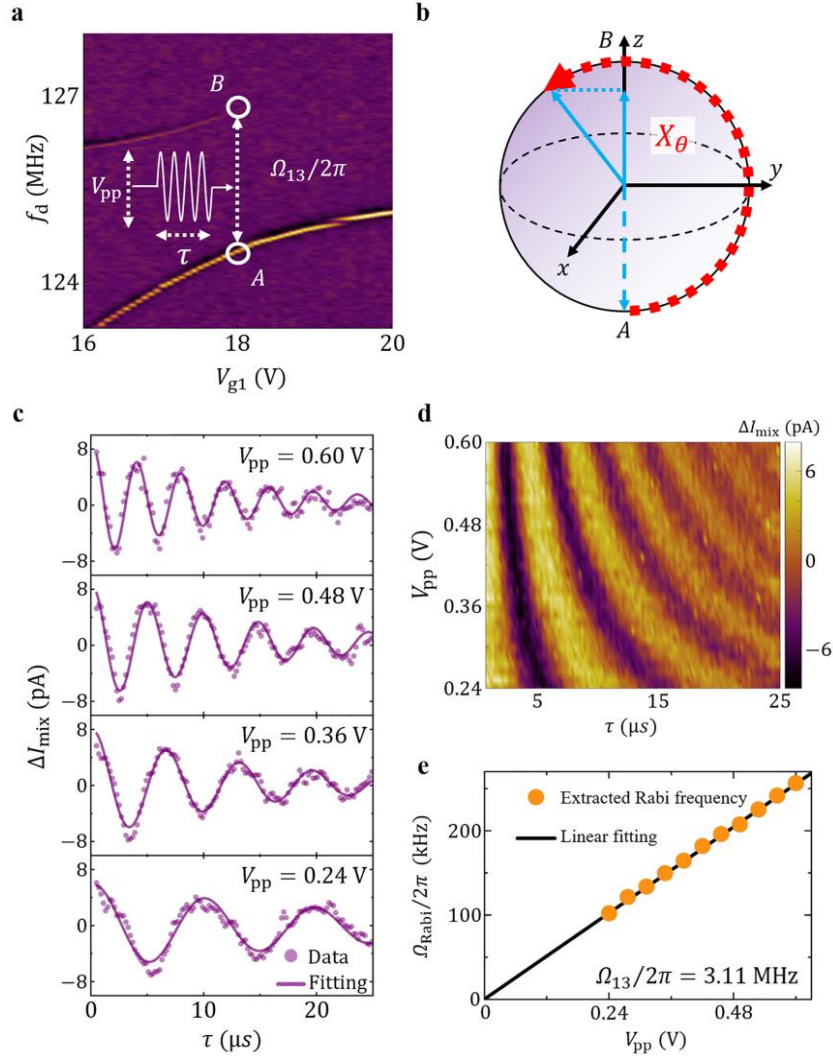

**Figure 2 | Phonon Rabi oscillations between spatially-separated resonators. a.** Spectrum of $R_1$ and $R_3$ at $V_{g2} = 17.5$ V and $V_{g3} = 22$ V with effective coupling strength $\Omega_{13}/2\pi = 3.11$ MHz. During the experiment, we first excite the coupled system to eigenmode $A$, then apply a microwave burst to generate rotations around the $x$-axis, and measure the mixing current at the frequency of eigenmode $A$. Each sequence is repeated for over $10^5$ times to gather the data for the average mixing current. **b.** Evolution of the system on the Bloch sphere. The dashed blue arrow marks the initial state. Under the microwave burst, the system rotates around the $x$-axis, following the red trajectory, and the burst duration $\tau$ determines the rotation angle. **c.**

Average mixing current as a function of burst duration with microwave amplitudes $V_{pp}$ = 0.60 V, 0.48 V, 0.36 V, and 0.24 V, respectively. The purple lines show the damped cosinusoidal fits to the experimental data. **d.** The Rabi oscillation versus the microwave amplitude $V_{pp}$. **e.** Extracted Rabi frequency $\Omega_{Rabi}/2\pi$ as a function of microwave amplitude in comparison with a linear fitting (black line).

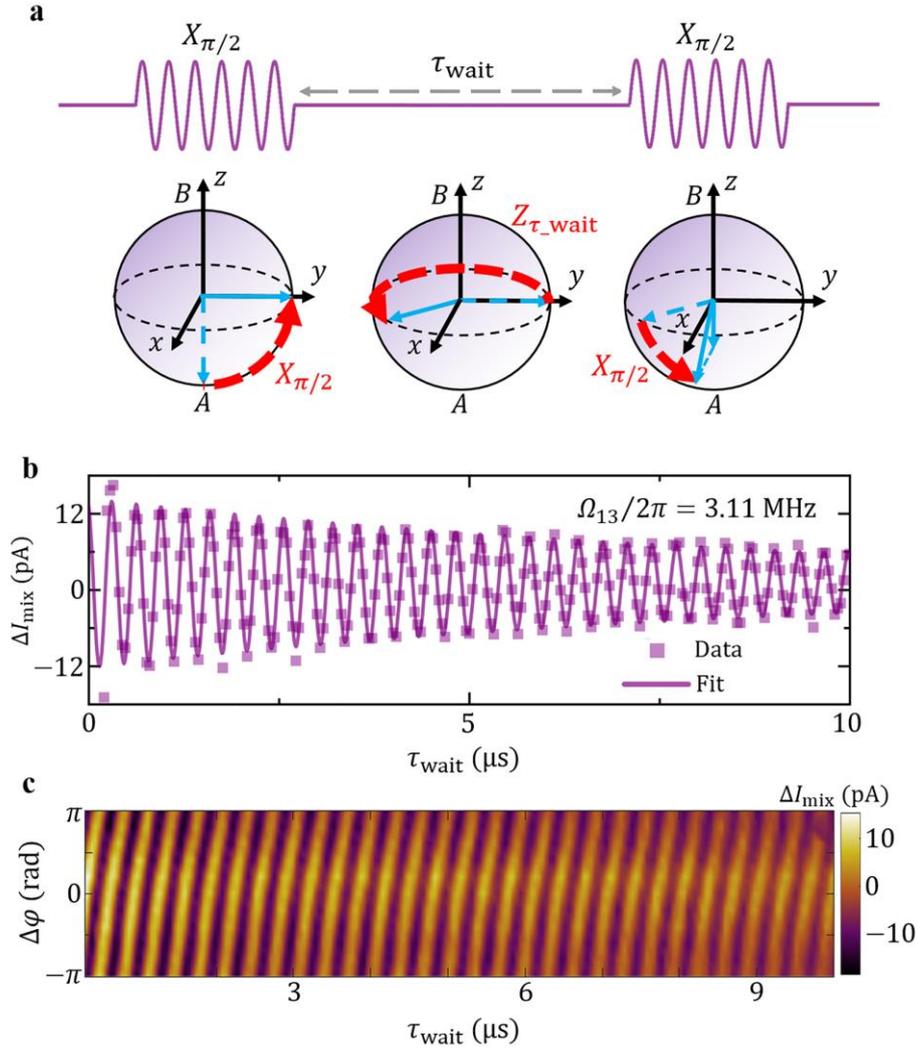

**Figure 3 | Phonon Ramsey interference between spatially-separated resonators. a.** Pulse sequence and system evolution in the Bloch sphere during Ramsey interference experiment. The pulse sequence consists of a $\pi/2$ rotation around the x-axis, a free evolution duration $\tau_{\text{wait}}$ and a second pulse around the *x*-axis (or a selected axis in the *x-y* plane). The dashed blue arrows indicate the initial state of this system. The dashed red arrows indicate the evolution trajectories. The solid blue arrows refer to the final state. **b.** Average mixing current as a function of $\tau_{\text{wait}}$ with coupling strength $\Omega_{13}/2\pi = 3.11$ MHz. The solid line is the numerical fitting using a damped cosinusoidal function, which gives an oscillation frequency $\Omega_{\text{Ramsey}}/2\pi = 3.12$ MHz,

in good agreement with the coupling strength. **c.** Ramsey interferences versus waiting time and relative phase $\Delta\varphi$ between the two $\pi/2$ pulses.

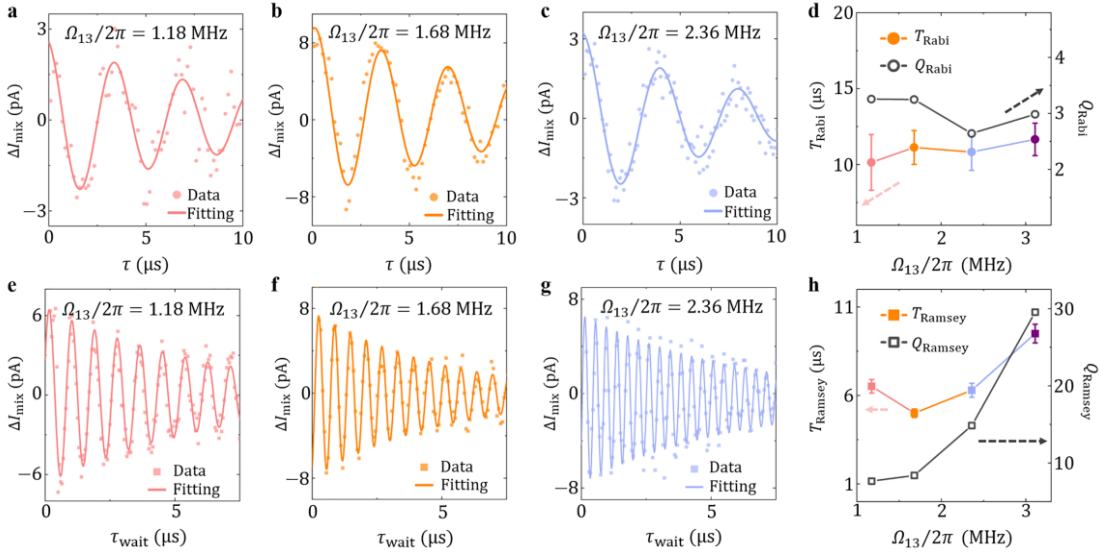

**Figure 4 | Rabi and Ramsey oscillations at different coupling strengths.** Phonon Rabi oscillations **(a-c)** and Ramsey interference **(e-g)** between spatially separated resonators at coupling strengths $\Omega_{13}/2\pi =$ 1.18 MHz, 1.68 MHz and 2.36 MHz, respectively. Solid lines are numerical fitting using damped cosinusoidal functions. **d.** Damping times of Rabi oscillations $T_{\text{Rabi}}$ and quality factor $Q_{\text{Rabi}} = T_{\text{Rabi}} \times \Omega_{\text{Rabi}}/2\pi$ as functions of coupling rate $\Omega_{13}/2\pi$. **h.** Damping times of Ramsey interference $T_{\text{Ramsey}}$ and quality factor $Q_{\text{Ramsey}} = T_{\text{Ramsey}} \times \Omega_{\text{Ramsey}}/2\pi$ as functions of coupling rate $\Omega_{13}/2\pi$. Error bars are given by fitting fidelities. Data points for $\Omega_{13}/2\pi =$ 3.11 MHz are taken from Figs. 2 and 3.